\documentclass[superscriptaddress,showpacs,preprintnumbers,amsmath,amssymb,prl,twocolumn]{revtex4}

\usepackage{graphicx}
\usepackage{dcolumn}
\usepackage{epsfig}
\usepackage{bm}

\def\eq#1{{(\ref{#1})}}

\newcommand{\beq}{\begin{equation}}
\newcommand{\eeq}{\end{equation}}
\newcommand{\ben}{\begin{eqnarray*}}
\newcommand{\een}{\end{eqnarray*}}
\begin{document}

\preprint{BNL--NT--04/21; June 9, 2004}

\title{Parity violation in hot QCD:\\ 

why it can happen, and how to look for it}

\author{Dmitri Kharzeev}
\affiliation{Physics Department,  \\ Brookhaven National Laboratory \\
Upton, NY 11973-5000}

\date{\today}

\begin{abstract}

The arguments for the possibility of violation of $\cal{P}$ and $\cal{CP}$ symmetries of 
strong interactions at finite temperature are presented. A new way of observing 
these effects in heavy ion collisions is proposed -- it is shown that 
 parity violation should manifest itself in the asymmetry between positive and 
 negative pions with respect to the reaction plane. Basing on topological 
 considerations, we derive a {\it lower} bound on the 
 magnitude of the expected asymmetry, which may appear within the reach of the current 
 and/or future heavy ion experiments.

\end{abstract}



\maketitle

The strong ${\cal CP}$ problem remains one of the most outstanding puzzles of the Standard Model. 
Even though several possible solutions have been put forward (for example, the axion scenario \cite{axion}),  
at present it is still not clear why ${\cal P}$ and ${\cal CP}$ invariances are respected by strong interactions.  

A few years ago, it was proposed that in the vicinity of the deconfinement phase 
transition QCD vacuum can possess metastable  domains leading to ${\cal P}$ and ${\cal CP}$ violation  
\cite{KPT}.   
It was also suggested that this phenomenon would manifest itself in specific 
correlations of pion momenta \cite{KPT,KP}. 
Such "${\cal P}$--odd bubbles" are a particular realization of an excited vacuum domain which may be produced in heavy ion collisions 
\cite{LeeWick}, and several other realizations have been proposed before \cite{Morley,DCC}. (For related studies of metastable vacuum states, especially in supersymmetric theories, see \cite{SUSY1,SUSY2,SUSY3}). However the peculiar pattern of ${\cal P}$ 
and ${\cal CP}$ breaking possessed by ${\cal P}$--odd bubbles may make them amenable to observation, as we will discuss in this letter.  

The existence of metastable ${\cal P}$--odd bubbles does not contradict the Vafa--Witten theorem \cite{VW} stating that 
${\cal P}$ and ${\cal CP}$ cannot be broken in the true ground state of QCD for $\theta =0$. Moreover, this theorem does not apply to QCD matter at finite isospin density \cite{SonMisha} and finite temperature \cite{Cohen}, where Lorentz--non-invariant ${\cal P}$--odd operators are allowed to have non-zero expectation values. Degenerate vacuum states with 
opposite parity were found \cite{PR} in the superconducting phase of QCD.
Parity broken phase also exists in lattice QCD with Wilson fermions \cite{Aoki}, but this phenomenon has been recognized as a lattice artifact for the case of mass--degenerate quarks; spontaneous ${\cal P}$ and ${\cal CP}$ breaking similar to the Dashen's phenomenon \cite{Dashen} can however occur for non--physical 
values of quark masses \cite{Creutz}.  ${\cal P}$--even, but ${\cal C}$--odd metastable states have also been argued to exist in hot gauge theories \cite{Chris}. The conditions for the applicability of Vafa-Witten theorem have been repeatedly re--examined in recent years \cite{VWth}. 
 
 Several dynamical scenarios for the decay of ${\cal P}$--odd bubbles have been considered \cite{decay}, and a numerical lattice calculation of 
 the fluctuations of topological charge in classical Yang--Mills fields has been performed \cite{Raju}.
 The studies of ${\cal P}$-- and ${\cal CP}$--odd correlations of pion momenta \cite{Voloshin,Sandweiss}, including those proposed in ref\cite{miklos}, have shown that such measurements are 
in principle feasible but would require large event samples. 
In addition, the magnitude of the expected effect despite the estimates done using the chiral Lagrangian approach \cite{KP} and a quasi-classical color field model \cite{DK} remained somewhat uncertain. 

In this letter, we will give additional arguments in favor of ${\cal P}$-- and ${\cal CP}$--breaking in a domain of a 
highly excited vacuum state. A new way of observing 
$\cal{P}$--odd effects in experiment through the asymmetry in the production of charged pions with respect 
to the reaction plane will then be proposed. It appears that the magnitude of the expected asymmetry 
can be estimated on the basis of topological considerations alone, and that the effect  
may be amenable to observation in the existing and/or future heavy ion experiments.

 Let us begin with a brief introduction to the strong ${\cal CP}$ problem.  
Strong interactions within the Standard Model are described by Quantum Chromo-Dynamics, with the Lagrangian 
\beq \label{qcd}
{\cal L} = -{1 \over 4} F^{\mu\nu}_{\alpha}F_{\alpha \mu\nu}  + \sum_f \bar{\psi}_f \left[ i \gamma^{\mu} 
(\partial_{\mu} - i g A_{\alpha \mu} t_{\alpha}) - m_f \right] 
\psi_f ,
\eeq
where $F^{\mu\nu}_{\alpha}$ and $A_{\alpha \mu}$ are the color field strength tensor and vector potential, respectively, 
$g$ is the strong coupling constant, $\psi_f$ are the quark fields of different flavors $f$ with masses $m_f$, and $t_{\alpha}$ the generators 
of the color $SU(3)$ group in the fundamental representation. The Lagrangian \eq{qcd} is symmetrical with 
respect to space parity  ${\cal P}$ and charge conjugation parity ${\cal C}$ transformations. 

However, these classical symmetries of QCD become questionable due to the interplay of quantum axial anomaly \cite{anomaly}
and classical topologically non-trivial solutions -- the instantons \cite{instanton}. The axial anomaly arises due to the fact that 
the renormalization of the theory \eq{qcd} cannot be performed in a chirally invariant way. As a result the flavor-singlet axial 
current $J_{\mu 5} = \bar{\psi}_f \gamma_{\mu} \gamma_5 \psi_f$ is no longer conserved even in the 
$m \to 0$ limit:
\beq \label{axanom}
 \partial^{\mu}  J_{\mu 5} = 2 m_f i \bar{\psi}_f \gamma_5 \psi_f - {N_f g^2 \over 16 \pi^2} F^{\mu\nu}_{\alpha} \tilde{F}_{\alpha \mu\nu}.
 \eeq
 where $\tilde{F}_{\alpha \mu\nu} = {1 \over 2} \epsilon_{\mu\nu\rho\sigma} F^{\alpha \rho\sigma}$. The 
 last term in \eq{axanom} is seemingly irrelevant since it can be written down as a full divergence, 
$
 F^{\mu\nu}_{\alpha}\tilde{F}_{\alpha \mu\nu} = \partial_{\mu} K^{\mu}
$, of the (gauge-dependent) topological gluon current  
$K^{\mu} = \epsilon^{\mu\nu\rho\sigma} A_{\alpha \nu} \left[F_{\alpha \rho\sigma} - {g \over 3} f_{\alpha\beta\gamma} 
A_{\beta \rho} A_{\gamma \sigma} \right]. 
$
However this conclusion is premature due to the existence of instantons which 
induce a change in the value of the chiral charge $Q_5 = \int d^3 x K^0$ associated with the topological current 
between $t = - \infty$ and $t= + \infty$: 
$
\nu = \int_{- \infty}^{+ \infty} dt {d Q_5 \over dt} = 2 N_f q[F],
$
where $q[F] = {\frac{g^2}{32 \pi^2}} \int d^4x F^{\mu\nu}_{\alpha} \tilde{F}_{\alpha \mu\nu}$ is the topological charge; 
for a one-instanton solution, $q = +1$. 

In the presence of degenerate topological vacuum sectors, an expectation value 
of an observable ${\cal O}$ has to be evaluated by first computing an average $\int_q D[\psi] D[\bar{\psi}] D[A] \exp(i S_{QCD}) {\cal O}(\psi, \bar{\psi}, A)$ over a sector with a fixed topological charge 
$q$, and then by summing over all sectors with the weight $f(q)$ \cite{Weinberg}.  The additivity constraint $f(q_1 + q_2) 
= f(q_1) f(q_2)$ restricts the weight to the form $f(q) = \exp(i \theta q)$, where $\theta$ is a free parameter. 
Recalling an explicit expression $q[F] = {\frac{g^2}{32 \pi^2}} \int d^4x F^{\mu\nu}_{\alpha} \tilde{F}_{\alpha \mu\nu}$ one can see that this procedure is equivalent to adding to the QCD Lagrangian  \eq{qcd} $S_{QCD} = \int d^4x {\cal L}_{QCD}$ a new term
\beq\label{theta}
{\cal L}_{\theta} = - {\theta \over 32 \pi^2}  g^2 F^{\mu\nu}_{\alpha} \tilde{F}_{\alpha \mu\nu}.
\eeq
Unless $\theta$ is identically equal to zero, ${\cal P}$ and ${\cal CP}$ invariances of QCD are lost.

One can eliminate the "$\theta$--term"  \eq{theta} (but not  ${\cal CP}$ violation itself) by a redefinition of the quark fields through 
the chiral rotation $\psi_f \to \exp( i \gamma_5 \theta_f/2)\ \psi_f$ with real phases $\theta = \sum_f \theta_f$. Indeed, because of the axial anomaly \eq{axanom}, this is equivalent to the replacement 
\beq\label{period}
\theta \to \theta + \sum_f \theta_f
\eeq
 so that the 
term \eq{theta} can be eliminated at the cost of introducing complex quark masses.  Introducing the left-- and right--handed quark fields 
$\psi_L = {1 \over 2} (1 - \gamma_5) \psi$,  $\psi_R = {1 \over 2} (1 + \gamma_5) \psi$, we can write the 
quark mass term of \eq{qcd} in the following form 
\beq \label{quark} 
{\cal L}_{quark} = -\sum_f \left( \hat{m}_f\  \bar{\psi}_{L, f} \psi_{R, f} + \hat{m}_f^* \ \bar{\psi}_{R, f} \psi_{L, f} \right)  ,
\eeq
where the real masses $m_f$ from the Lagrangian \eq{qcd} have been replaced by complex mass parameters 
$\hat{m}_f = m_f \exp(i \theta_f)$. Because of \eq{period}, all $\cal{CP}$-violating phase can be attributed to a 
single quark flavor, say $u$, so that $\theta = \theta_u$, $\theta_d = \theta_s = 0$. Therefore if at least one of the 
quarks is massless, the $\cal{CP}$-violating phase would not have any observable effect. 
From now on, we will rotate for simplicity all $\cal{CP}$ violating phase into the "up" quark  of mass $m \equiv m_u$; this does not lead to any loss of generality. 
We would like to emphasize again that quark masses are absolutely essential in the strong $\cal{CP}$ violation -- this will be  important in what follows.    

The complex mass parameters in \eq{quark} can be treated as "spurion" fields \cite{tHooft}, with an insertion of $\hat{m}$ flipping left quarks 
into right, and vice versa for $\hat{m}^*$. This "spurion" field is associated with a canonical chiral charge operator
\beq \label{chircharge}
\Delta Q_5 = 2 \left( \hat{m}^* {\partial \over \partial \hat{m}^*} -  \hat{m} {\partial \over \partial \hat{m}} \right) = 
2 i {\partial \over \partial \theta}. 
\eeq
The parity-odd effect of the complex mass parameters inducing the difference between the left- and right-handed fermions can be made completely manifest by re-writing the $u$ quark part of \eq{quark} as 
\beq\label{lr}
{\cal L}_{\theta} = - m \cos \theta \left(  \bar{u}_L u_R + \bar{u}_R u_L \right) - i m \sin \theta  \left(  \bar{u}_L u_R -  \bar{u}_R u_L \right).
\eeq
Parity violation in strong interactions has been never detected, and stringent limits exist 
on the value of $\cal{CP}$ violating phase $\theta < 3\times 10^{-10}$. 
This means that in the physical vacuum the 
"spurion" field $\hat{m} = m \exp(i \theta)$ has a real expectation value determined 
by the quark masses $\left< \hat{m} \right> = m$. Because $\hat{m}$ and $\theta$ cannot have any space-time 
dependence in the physical vacuum, the "spurion" field does not carry any energy or momentum. 

The metastable $\cal{P}$ and $\cal{CP}$ odd state of ref \cite{KPT} 
acts as a localized in space and time vacuum domain with $\theta = \theta( {\bf{x}}, t ) \neq 0$;  the space-time dependence of $\theta$ and thus of 
$\hat{m}( {\bf{x}}, t ) = m \exp(i \theta( {\bf{x}}, t ))$ implies that the chiral charge operator \eq{chircharge} no longer commutes with the 
operator of momentum and the Hamiltonian. 
Therefore the field $\hat{m}$ can now scatter quarks and create quark--antiquark pairs with 
non--zero chirality. What is the definition of chirality in this situation? This question is not trivial since as we have 
seen above parity violation in QCD is possible only if all quark masses are different from zero, and the 
definition of chirality for a massive fermion is not Lorentz invariant and depends on the frame.

Let us discuss this in more detail. 
Consider the second term in \eq{lr} which is responsible for parity violation; in terms 
of the two--component spinors $\chi$ and Pauli spin matrices ${\bf{\sigma}}$ it involves
\beq \label{nonrel}
\chi^+ {\bf \sigma} ({\bf n - n'}) \chi, 
\eeq  
where    $ {\bf n} = {\bf p} / p$  is the unit vector in the direction of the quark momentum ${\bf p}$, and we have 
assumed that the quark energy $E \gg m$. In the vacuum, the "spurion" field $\hat{m}$ carries no energy or momentum, 
so the interaction of quarks with spurions leaves ${\bf p} = {\bf p}'$, ${\bf n} = {\bf n}'$. This means that the 
chirality change is possible only through the flip of the spin of the quark, which changes the sign of the 
spin projection on the momentum, so that $\left< {\bf \sigma n} \right>_i  = - \left< {\bf \sigma n} \right>_f$.

Consider now a domain of excited QCD vacuum with $\theta = \theta( {\bf{x}}, t )$; the "spurion" field associated 
with it can now transfer energy and 
momentum to the quarks, so that ${\bf p} \neq {\bf p}'$ in the quark-spurion interaction vertex. 
Moreover, the rest frame of the domain defines a preferred reference frame in which the chirality of the massive quark  
is to be measured. 
If the domain is axially symmetric, and $\theta = \theta(r,\Omega)$ depends only on the polar angle $\Omega$ and not on the azimuthal angle $\phi$ (which as we will soon see is the case for QCD matter produced in heavy ion collisions), 
this symmetry by Wigner-Eckart theorem defines the appropriate quantization axis for the quark spin ${\bf \sigma}$. 
{\it Such a domain can generate chirality not by flipping the spins of the quarks, but by inducing up--down asymmetry 
(as measured with respect to the symmetry axis) 
in the production of quarks and antiquarks.} 

Formally, this happens because the operator of chiral charge \eq{chircharge}, corresponding to the rotation in the 
$\theta$ space, in this case commutes with the operator of rotations $-i {\partial \over \partial \phi}$ in azimuthal 
angle, but not with rotations in polar angle $\Omega$.  
If the spins of the quark and antiquark are aligned parallel to the 
symmetry axis of the domain, "right" quark would refer to the quark emitted in the upper hemisphere (along the 
direction of the symmetry axis, with ${\bf \sigma n} > 0$), and viceversa for the "left" antiquark. Therefore, 
a domain   with $\theta = \theta( \bf{x}, t )$ can generate {\it spatial} asymmetry in the production of $\bar{u}u$ and 
other quark pairs. 
In terms of the observable charged pions, this would mean that positive and negative pions will be produced 
asymmetrically with respect to the symmetry axis.  
Because of the overall charge conservation, this implies that there will be more positive than negative pions in the upper hemisphere, and more negative than positive pions in the lower hemisphere (the sign of the asymmetry 
is of course determined by the sign of the topological chiral charge of the domain). 

The spatial separation of positive and negative 
charges will induce an electric dipole moment (e.d.m.) in the system, which is a clear signature of ${\cal CP}$ violation. Searching for  
the fluctuations of $\theta$ angle through the spatial separation of electric charges in the hot quark--gluon fireball is analogous to 
the proposal of constraining the value of  $\theta$ in the vacuum by measuring the e.d.m. of the neutron \cite{edm}.
In the framework of the chiral lagrangean description \cite{edm}, the spatial asymmetry of the pion cloud around the neutron is caused 
by the ${\cal P}$--odd $\pi N$ coupling. Recently, the phenomenon of the spatial separation of quarks with different electric charges 
at finite $\theta$ has also been demonstrated in the framework of the instanton liquid model \cite{Faccioli:2004ys}.

Would a $\theta$ domain produced in a heavy ion collision have a symmetry axis? Consider 
two symmetrical heavy ions with mass number $A$ colliding with the center-of-mass energy $\sqrt{s}$ per nucleon pair, at an impact parameter 
$b$. In the c.m.s. frame the initial angular momentum of this system is $L \approx A | [{\bf  b} \times {\bf p}] | \simeq A\ b\ \sqrt{s}/2$. With $\sqrt{s} = 200$ GeV 
(the energy of the RHIC collider), we have $L \simeq A/2  \ b[{\rm fm}] \times 10^3$  units of angular momentum in the system.
After the collision, part of this angular momentum is carried away from the produced fireball by the "spectator" nucleons, 
but it is clear that the produced matter must have thousands of units of angular momentum. 
This angular momentum is pointing perpendicular to the reaction plane, which can be reconstructed both by 
detecting the directions of forward fragments in the fragmentation regions on both sides, and by studying the particle correlations at mid-rapidity region. The angular momentum vector provides us with the symmetry axis discussed above. 
Moreover, we can now supplement our arguments with a simple semi--classical 
picture: rotating deconfined color charges generate chromo-magnetic field ${\bf H}$ parallel to the angular momentum vector, and the quarks spins align along ${\bf H}$.   

What is the magnitude of the expected effect? Fortunately we can estimate it without invoking any models for the $\cal{CP}$--odd domain structure. 
Let us choose the polar axis along the vector of angular momentum; the distrubution $N_+$ ($N_-$) of the produced $u$ ($\bar{u}$) quarks in the polar angle $\Omega$ according to \eq{lr},\eq{nonrel} will 
then be given by
\beq \label{distr}
{dN_{\pm} \over d  \Omega} = {\rm const} \ \left(1 \pm \kappa \ \cos \Omega \right) \ \sin \Omega.
\eeq
As usual, the ${\cal CP}$--odd term in \eq{distr}  appears due to the interference of $\cal{CP}$ breaking term \eq{nonrel} with the 
$\cal{CP}$ even terms. Because of this, and because most of the quarks will be produced by parity--conserving interactions, one cannot 
evaluate the constant $\kappa$ in \eq{distr} from \eq{lr} alone. Moreover, the dynamics of the collision will severely affect the shape of the distribution, adding parity--even harmonics to \eq{distr}. Nevertheless,  since \eq{lr} is the only source of parity violation and all other interactions conserve parity, the up--down asymmetry in the production of $u$ quarks defined as
\beq \label{asym}
A_u =  { N_R - N_L \over N_R + N_L} = 
\eeq
\beq \nonumber
=  \left( \int_{0}^{\pi} {dN_{+} \over d  \Omega} \right)^{-1} \  \left( {\int_0^{\pi/2} {dN_{+} \over d  \Omega}  - \int_{\pi/2}^{\pi} {dN_{+} \over d  \Omega}} \right) ,
\eeq
will be preserved in the subsequent evolution of the system. Obviously, the asymmetry for $\bar{u}$ antiquarks will be $A_{\bar{u}} = - A_u = -\kappa/2$.
The asymmetry between $u$ and $\bar{u}$ quarks \eq{asym} is not directly observable; 
however if the hadronization process preserves ${\cal P}$ and ${\cal CP}$, it should translate into the observable asymmetry in the production 
of charged pions; we will thus assume that $ A_{\pi^+} = - A_{\pi^-} = A_u $. 

Let us consider a ${\cal P}$--odd domain with a topological charge $Q \geq 1$. Then $N_R - N_L = Q$ in \eq{asym}; if the total 
multiplicity of positive pions is $N_R + N_L = N_{\pi^+}$ we get for the asymmetry an estimate
\beq \label{est}
A_{\pi^+} = - A_{\pi^-} \simeq {Q \over N_{\pi^+}},
\eeq 
where $Q \geq 1$. It is important to note that topological charge $Q$ of the domain is a conserved quantity, whereas the 
multiplicity of final state pions $N_{\pi}$ strongly fluctuates. In the deconfined phase, the probability of forming topologically charged 
domains is not suppressed so one may expect the $\cal{CP}$--odd effects in almost every heavy ion collision event at sufficiently 
high energy.

Soft particles produced in high-energy collisions are known to be correlated 
over about one unit of rapidity, which would most likely be a typical extent of a ${\cal P}$--odd bubble in rapidity space, so one can take $N_{\pi^+} = dn_{\pi^+}/dy$. Even in the central rapidity region of heavy ion collisions the multiplicity of positive pions can slightly exceed the one for negative pions because the colliding 
nuclei are positively charged; however the normalized asymmetries \eq{asym} of course should still be equal and opposite in sign. 
(If the temperature is low and the isospin asymmetry is large, ${\cal P}$--odd condensates can form in the system \cite{SonMisha}, 
but these conditions are not met in heavy ion collisions).

 The multiplicity $dn_{\pi^+}/dy$ depends 
on the centrality of the collision (apart from the energy and the mass number of the colliding ions); very peripheral collisions are most likely 
incapable of producing a sufficiently extended volume of hot matter, so excluding them the multiplicity per unit of rapidity in RHIC $Au-Au$ events typically varies within the limits $100 \leq N_{\pi^+} \leq 300$. The expected magnitude of the asymmetry \eq{est} is thus $A_{\pi^+} \sim 10^{-2}$. 
It may be possible to detect asymmetry of this magnitude by studying $\pi^+\pi^+$ and  $\pi^-\pi^-$ correlations with respect to the reaction plane 
of the collision. The average angle $\delta \chi = \pi/2 - \Omega$ of $\pi^+$ meson with respect to the reaction plane according to \eq{distr} is 
$\left< \delta \chi_{\pi^+} \right> = 2 \kappa / 3 = 4 A_{\pi^+} / 3 \sim 10^{-2}$. 
While the parity violation of that magnitude may well be amenable to observation, an experimental study of the effect will 
require an ingenious high--precision method of correlating pion momentum asymmetries with the reaction plane, reconstructed from the 
elliptic flow and/or from the directions of the forward fragments. 

The ideas of using a decay of an oriented system to test fundamental symmetries date back 
to  the work  \cite{LeeYang} which led to the discovery of parity violation in weak interactions.
The spatial separation of positive $u$ quarks and negative $\bar{u}$ anti-quarks in hot QCD matter (and the resulting spatial asymmetry for $\pi^+$ and $\pi^-$ production) 
induces an electric dipole moment of the system. 

An observation of such an asymmetry in heavy ion collisions would signal for the 
first time the possibility of ${\cal P}$ and ${\cal CP}$--odd effects in strong interactions.  Moreover, since the QCD vacuum is 
known to conserve parity, such an observation would establish unambiguously the creation of a different phase of quark--gluon matter.

\vskip0,2cm
 
\begin{acknowledgments}

I am indebted to T.D. Lee, J. Sandweiss and S.Voloshin for stimulating and enlightening discussions. 
Useful conversations with P.Bond, D.Budker, M.Creutz, M.Pospelov and A. Ritz are also gratefully acknowledged.  
This work was supported
by the U.S. Department of Energy under Contract No. DE-AC02-98CH10886.  

\end{acknowledgments}


\end{document}